%% file: main.tex
\documentclass[sigconf]{acmart}
\pdfoutput=1
\usepackage{booktabs} 
\usepackage{graphicx}
\usepackage{svg}

\input{header}

\setcopyright{acmcopyright}

\usepackage{lmodern}

\acmDOI{10.475/123_4}

\acmISBN{123-4567-24-567/08/06}

\acmConference[MAL-IOT19]{Malicious Software and Hardware in Internet of Things}{May 2019}{Alghero, Italy}
\acmYear{2019}
\copyrightyear{2019}

\acmArticle{4}
\acmPrice{15.00}

\begin{document}
	
	\title{Highway to HAL}
	\subtitle{Open-Sourcing the First Extendable Gate-Level Netlist Reverse Engineering Framework}
	\author{Sebastian Wallat*, Nils Albartus, Steffen Becker, Max Hoffmann, Maik Ender, Marc Fyrbiak, Adrian Drees, Sebastian Maaßen, Christof Paar}
	\authornote{Sebastian Wallat and Christof Paar are also affiliated with University of Massachussets, Amherst, USA}
	\affiliation{%
		\institution{\textit{Ruhr-Universität Bochum}}
		\institution{\textit{Horst Görtz Institut für IT-Security}}
		\streetaddress{Universitätsstr. 150, 44801 Bochum}
		\city{Bochum}
		\state{Germany}
		\postcode{44801}
	}
	\email{{firstname.lastname}@rub.de}

	\renewcommand{\shortauthors}{S. Wallat et al.}
	
	%
	\begin{abstract}
	    Since hardware oftentimes serves as the root of trust in our modern interconnected world, malicious hardware manipulations constitute a ubiquitous threat in the context of the \ac{IOT}.
	    Hardware reverse engineering is a prevalent technique to detect such manipulations.
	    
	    Over the last years, an active research community has significantly advanced the field of hardware reverse engineering.
	    Notably, many open research questions regarding the extraction of functionally correct netlists from \acp{FPGA} or \acp{ASIC} have been tackled.
	    In order to facilitate further analysis of recovered netlists, a software framework is required, serving as the foundation for specialized algorithms.
	    Currently, no such framework is publicly available.
	    
	    Therefore, we provide the first open-source gate-library agnostic framework for gate-level netlist analysis. In this positional paper, we demonstrate the workflow of our modular framework \HAL on the basis of two case studies and provide profound insights on its technical foundations.
	\end{abstract}

	\maketitle
	
	
	\input{sections/introduction}
	
	
	\input{sections/technical_background}

	
	\input{sections/hal}

	
	\input{sections/case-studies}

	\input{sections/conclusion}

	\section*{Acknowledgment}
	The research was supported in part by ERC Advanced Grant 695022 and NSF award NS-1563829.

	\bibliographystyle{ACM-Reference-Format}
	\bibliography{bibliography}
	
\end{document}

%% file: header.tex
\usepackage[utf8]{inputenc}
\usepackage[english]{babel}

\usepackage{url}
\usepackage{hyperref} 

\usepackage{amsmath,amssymb,amsthm,wasysym,mathtools,esvect}

\usepackage[babel=true]{csquotes}
\usepackage{cleveref}

\usepackage{amsmath,amssymb,amsthm,wasysym,mathtools,esvect,dsfont}

\usepackage{graphicx,color,xcolor}

\usepackage{tikz}
\usetikzlibrary{automata,arrows,calc,shadows.blur,shapes.symbols}
\usepackage{circuitikz}
\usepackage{pgf}
\usepackage{pgfplots}
\usepackage[most]{tcolorbox}

\definecolor{persianrose}{rgb}{1.0, 0.16, 0.64}
\definecolor{goldenyellow}{rgb}{1.0, 0.87, 0.0}

\makeatother

\usepackage{float}

\usepackage{tikz}
\usetikzlibrary{automata}
\usetikzlibrary{arrows}
\usetikzlibrary{shadows}
\usetikzlibrary{decorations.text}
\usetikzlibrary{circuits.logic.US} 
\usetikzlibrary{matrix}

\usepackage{pgfplots}

\usepackage{enumitem}

\usepackage{environ}

\usepackage{tabularx}
\usepackage{makecell}
\usepackage{multirow}
\usepackage{colortbl}
\usepackage{collcell}
\usepackage{threeparttable}
\usepackage{adjustbox}
\usepackage{booktabs}

\usepackage{listings}
\let\origthelstnumber\thelstnumber
\makeatletter
\newcommand*\Suppressnumber{%
  \lst@AddToHook{OnNewLine}{%
    \let\thelstnumber\relax%
     \advance\c@lstnumber-\@ne\relax%
    }
}

\newcommand*\Reactivatenumber[1]{%
  \setcounter{lstnumber}{\numexpr#1-1\relax}
  \lst@AddToHook{OnNewLine}{%
   \let\thelstnumber\origthelstnumber%
   \refstepcounter{lstnumber}
  }
}

\makeatother
\lstdefinestyle{vhdl_style}
{
    language=VHDL,
    float=!htb,
    basicstyle=\ttfamily\footnotesize,
    identifierstyle=\bfseries\color{black},
    keywordstyle=\bfseries\color{persianrose},
    stringstyle=\bfseries\color{yellow},
    commentstyle=\bfseries\color{gray},
    columns=flexible,
    frame=single,
    showspaces=false,
    showstringspaces=false,
    numberstyle=\tiny,
    stepnumber=1,
    breaklines=true,
    xrightmargin=-\fboxsep,
    backgroundcolor=\color{white},
    captionpos=t,
    mathescape,
    escapechar=\%
}

\usepackage{algorithm}
\usepackage[noend]{algpseudocode}
\algrenewcommand\algorithmicrequire{\textbf{\ \ Input:}}
\algrenewcommand\algorithmicensure{\textbf{Output:}}

\usepackage{microtype}
\usepackage{xspace}
\usepackage{pifont}
\usepackage{comment}

\usepackage{ulem}

\usepackage[nolist]{acronym}
\input{acronyms.tex}

\definecolor{lightcoral}{rgb}{0.94, 0.5, 0.5}
\definecolor{lightskyblue}{rgb}{0.53, 0.81, 0.98}

\newcommand{\CC}[1][]{$\text{C\hspace{-.25ex}}^{_{_{_{++}}}}\ifthenelse{\equal{#1}{}}{}{\text{\hspace{-.625ex}#1}}$\xspace}

\newcommand{\circled}[2][]{
  \tikz[baseline=(char.base)]{
    \node[shape=circle,draw,inner sep=1pt,fill=black]
    (char) {\phantom{\ifblank{#1}{#2}{#1}}};
    \node[text=white] at (char.center) {\makebox[0pt][c]{\textbf#2}};}\xspace}
\robustify{\circled}

\makeatletter
\newcommand{\removeifnextchar}[2]{%
    \begingroup
    \ltx@LocToksA{\endgroup#2}%
    \ltx@ifnextchar@nospace{#1}{%
        \def\next{\the\ltx@LocToksA}%
        \afterassignment\next
        \let\scratch= %
    }{%
        \the\ltx@LocToksA
    }%
}

\newcommand{\etal}{\protect\removeifnextchar{.}{et~al.\xspace}}

\makeatother

\newcommand{\romannumber}[1]{\uppercase\expandafter{\romannumeral#1}}

\ifdefined\algorithmautorefname

\else
  \newcommand{\algorithmautorefname}{Algorithm}
\fi 

\ifdefined\definitionautorefname

\else 
  \newcommand{\definitionautorefname}{Definition}
\fi

\addto\extrasenglish{  
  \ifdefined\algorithmautorefname
    \renewcommand{\algorithmautorefname}{Algorithm}
  \fi 
  \ifdefined\definitionautorefname
    \renewcommand{\definitionautorefname}{Definition}
  \fi

}

\newcommand{\Figure}[1]{\textcolor{blue}{\autoref{#1}}}

\newcommand{\Section}[1]{\textcolor{blue}{\autoref{#1}}}

\renewcommand{\Figure}[1]{\autoref{#1}}

\renewcommand{\Section}[1]{\autoref{#1}}

\newcommand{\HAL}{\textnormal{\textsf{HAL}}\xspace}

\newcommand*{\MinNumber}{0.0}
\newcommand*{\MidNumber}{0.76}
\newcommand*{\MaxNumber}{1.0}

\newcommand{\ApplyGradient}[1]{
  \ifdim #1 pt > \MidNumber pt
    \pgfmathsetmacro{\PercentColor}{max(min(100.0*(#1 - \MidNumber)/(\MaxNumber-\MidNumber),100.0),0.00)} %
    \colorbox{green!\PercentColor!yellow}{#1}
  \else
    \pgfmathsetmacro{\PercentColor}{max(min(100.0*(\MidNumber - #1)/(\MidNumber-\MinNumber),100.0),0.00)} %
    \colorbox{red!\PercentColor!yellow}{#1}
  \fi
}

\newcolumntype{R}{>{\collectcell\ApplyGradient}c<{\endcollectcell}}

\makeatletter
\pgfdeclareshape{dff}{
	\savedanchor\northeast{%
		\pgfmathsetlength\pgf@x{\pgfshapeminwidth}%
		\pgfmathsetlength\pgf@y{\pgfshapeminheight}%
		\pgf@x=0.5\pgf@x
		\pgf@y=0.5\pgf@y
	}
	\savedanchor\southwest{%
		\pgfmathsetlength\pgf@x{\pgfshapeminwidth}%
		\pgfmathsetlength\pgf@y{\pgfshapeminheight}%
		\pgf@x=-0.5\pgf@x
		\pgf@y=-0.5\pgf@y
	}
	\inheritanchorborder[from=rectangle]
	
	\anchor{center}{\pgfpointorigin}
	\anchor{north}{\northeast \pgf@x=0pt}
	\anchor{east}{\northeast \pgf@y=0pt}
	\anchor{south}{\southwest \pgf@x=0pt}
	\anchor{west}{\southwest \pgf@y=0pt}
	\anchor{north east}{\northeast}
	\anchor{north west}{\northeast \pgf@x=-\pgf@x}
	\anchor{south west}{\southwest}
	\anchor{south east}{\southwest \pgf@x=-\pgf@x}
	\anchor{text}{
		\pgfpointorigin
		\advance\pgf@x by -.5\wd\pgfnodeparttextbox%
		\advance\pgf@y by -.5\ht\pgfnodeparttextbox%
		\advance\pgf@y by +.5\dp\pgfnodeparttextbox%
	}
	
	\anchor{D}{
		\pgf@process{\northeast}%
		\pgf@x=-1\pgf@x%
		\pgf@y=.6\pgf@y%
	}
	\anchor{CLK}{
		\pgf@process{\northeast}%
		\pgf@x=-1\pgf@x%
		\pgf@y=-.7\pgf@y%
	}
	\anchor{Q}{
		\pgf@process{\northeast}%
		\pgf@y=.6\pgf@y%
	}
	\backgroundpath{
		\pgfpathrectanglecorners{\southwest}{\northeast}
		\pgf@anchor@dff@CLK
		\pgf@xa=\pgf@x \pgf@ya=\pgf@y
		\pgf@xb=\pgf@x \pgf@yb=\pgf@y
		\pgf@xc=\pgf@x \pgf@yc=\pgf@y
		\pgfmathsetlength\pgf@x{1.6ex} 
		\advance\pgf@ya by \pgf@x
		\advance\pgf@xb by \pgf@x
		\advance\pgf@yc by -\pgf@x
		\pgfpathmoveto{\pgfpoint{\pgf@xa}{\pgf@ya}}
		\pgfpathlineto{\pgfpoint{\pgf@xb}{\pgf@yb}}
		\pgfpathlineto{\pgfpoint{\pgf@xc}{\pgf@yc}}
		\pgfclosepath
		
		\begingroup
		\tikzset{flip flop/port labels} 
		\tikz@textfont
		
		\pgf@anchor@dff@D
		\pgftext[left,base,at={\pgfpoint{\pgf@x}{\pgf@y}},x=\pgfshapeinnerxsep]{\raisebox{-0.75ex}{\textcolor{black}{D}}}
		
		\pgf@anchor@dff@Q
		\pgftext[right,base,at={\pgfpoint{\pgf@x}{\pgf@y}},x=-\pgfshapeinnerxsep]{\raisebox{-.75ex}{Q}}
		
		\endgroup
	}
}

\tikzset{add font/.code={\expandafter\def\expandafter\tikz@textfont\expandafter{\tikz@textfont#1}}}

\tikzset{flip flop/port labels/.style={font=\sffamily\scriptsize}}
\tikzset{every dff node/.style={draw,minimum width=2cm,minimum
		height=2.5cm,very thick,inner sep=1mm,outer sep=0pt,cap=round,add
		font=\sffamily}}

\pgfdeclareshape{lut}{
	\savedanchor\northeast{%
		\pgfmathsetlength\pgf@x{\pgfshapeminwidth}%
		\pgfmathsetlength\pgf@y{\pgfshapeminheight}%
		\pgf@x=0.5\pgf@x
		\pgf@y=0.5\pgf@y
	}
	\savedanchor\southwest{%
		\pgfmathsetlength\pgf@x{\pgfshapeminwidth}%
		\pgfmathsetlength\pgf@y{\pgfshapeminheight}%
		\pgf@x=-0.5\pgf@x
		\pgf@y=-0.5\pgf@y
	}
	\inheritanchorborder[from=rectangle]
	
	\anchor{center}{\pgfpointorigin}
	\anchor{north}{\northeast \pgf@x=0pt}
	\anchor{east}{\northeast \pgf@y=0pt}
	\anchor{south}{\southwest \pgf@x=0pt}
	\anchor{west}{\southwest \pgf@y=0pt}
	\anchor{north east}{\northeast}
	\anchor{north west}{\northeast \pgf@x=-\pgf@x}
	\anchor{south west}{\southwest}
	\anchor{south east}{\southwest \pgf@x=-\pgf@x}
	\anchor{text}{
		\pgfpointorigin
		\advance\pgf@x by -.5\wd\pgfnodeparttextbox%
		\advance\pgf@y by -.5\ht\pgfnodeparttextbox%
		\advance\pgf@y by +.5\dp\pgfnodeparttextbox%
	}
	
	\anchor{IONE}{
		\pgf@process{\northeast}%
		\pgf@x=-1\pgf@x%
		\pgf@y=.5\pgf@y%
	}
	\anchor{ITWO}{
		\pgf@process{\northeast}%
		\pgf@x=-1\pgf@x%
		\pgf@y=0\pgf@y%
	}
	\anchor{ITHREE}{
		\pgf@process{\northeast}%
		\pgf@x=-1\pgf@x%
		\pgf@y=-.5\pgf@y%
	}
	\anchor{Q}{
		\pgf@process{\northeast}%
		\pgf@y=.5\pgf@y%
	}
	\backgroundpath{
		\pgfpathrectanglecorners{\southwest}{\northeast}
		
		\begingroup
		\tikzset{lut/port labels} 
		\tikz@textfont
		
		\pgf@anchor@lut@IONE
		\pgftext[left,base,at={\pgfpoint{\pgf@x}{\pgf@y}},x=\pgfshapeinnerxsep]{\raisebox{-0.75ex}{I0}}
		
		\pgf@anchor@lut@ITWO
		\pgftext[left,base,at={\pgfpoint{\pgf@x}{\pgf@y}},x=\pgfshapeinnerxsep]{\raisebox{-0.75ex}{I1}}
		
		\pgf@anchor@lut@ITHREE
		\pgftext[left,base,at={\pgfpoint{\pgf@x}{\pgf@y}},x=\pgfshapeinnerxsep]{\raisebox{-0.75ex}{I2}}
		
		\pgf@anchor@lut@Q
		\pgftext[right,base,at={\pgfpoint{\pgf@x}{\pgf@y}},x=-\pgfshapeinnerxsep]{\raisebox{-.75ex}{O}}
		\endgroup
	}
}

\tikzset{add font/.code={\expandafter\def\expandafter\tikz@textfont\expandafter{\tikz@textfont#1}}}

\tikzset{lut/port labels/.style={font=\sffamily\scriptsize}}
\tikzset{every lut node/.style={draw,minimum width=2cm,minimum
		height=3cm,very thick,inner sep=1mm,outer sep=0pt,cap=round,add
		font=\sffamily}}
\makeatother

%% file: acronyms.tex
\begin{acronym}

\acro{3DES}{Triple-DES}

\acro{AES}{Advanced Encryption Standard}
\acro{ALU}{Arithmetic Logic Unit}
\acro{API}{Application Programming Interface}
\acro{ARX}{Addition Rotation XOR}
\acro{ASIC}{Application Specific Integrated Circuit}
\acro{ASIP}{Application Specific Instruction-Set Processor}
\acro{AS}{Active Serial}

\acro{BDD}{Binary Decision Diagram}
\acro{BGL}{Boost Graph Library}
\acro{BNF}{Backus-Naur Form}
\acro{BRAM}{Block-Ram}

\acro{CBC}{Cipher Block Chaining}
\acro{CFB}{Cipher Feedback Mode}
\acro{CFG}{Control Flow Graph}
\acro{CLB}{Configurable Logic Block}
\acro{CLI}{Command Line Interface}
\acro{COFF}{Common Object File Format}
\acro{CPA}{Correlation Power Analysis}
\acro{CPU}{Central Processing Unit}
\acro{CRC}{Cyclic Redundancy Check}
\acro{CTR}{Counter}

\acro{DC}{Direct Current}
\acro{DES}{Data Encryption Standard}
\acro{DFA}{Differential Frequency Analysis}
\acro{DFT}{Discrete Fourier Transform}
\acro{DLL}{Dynamic Link Library}
\acro{DMA}{Direct Memory Access}
\acro{DNF}{Disjunctive Normal Form}
\acro{DPA}{Differential Power Analysis}
\acro{DSO}{Digital Storage Oscilloscope}
\acro{DSP}{Digital Signal Processing}
\acro{DUT}{Design Under Test}

\acro{ECB}{Electronic Code Book}
\acro{ECC}{Elliptic Curve Cryptography}
\acro{EEPROM}{Electrically Erasable Programmable Read-only Memory}
\acro{EMA}{Electromagnetic Emanation}
\acro{EM}{electro-magnetic}

\acro{FFT}{Fast Fourier Transformation}
\acro{FF}{Flip Flop}
\acro{FI}{Fault Injection}
\acro{FIR}{Finite Impulse Response}
\acro{FPGA}{Field Programmable Gate Array}
\acro{FSM}{Finite State Machine}
\acro{FIB}{Focused Ion Beam}

\acro{GUI}{Graphical User Interface}

\acro{HDL}{Hardware Description Language}
\acro{HD}{Hamming Distance}
\acro{HF}{High Frequency}
\acro{HSM}{Hardware Security Module}
\acro{HW}{Hamming Weight}

\acro{IC}{Integrated Circuit}
\acro{IO}[I/O]{Input/Output}
\acro{IOB}{Input Output Block}
\acro{IOT}[IoT]{Internet of Things}
\acro{IP}{Intellectual Property}
\acro{ISA}{Instruction Set Architecture}
\acro{IV}{Initialization Vector}

\acro{JTAG}{Joint Test Action Group}

\acro{KAT}{Known Answer Test}

\acro{LFSR}{Linear Feedback Shift Register}
\acro{LSB}{Least Significant Bit}
\acro{LUT}{Look-up table}

\acro{MAC}{Message Authentication Code}
\acro{MIPS}{Microprocessor without Interlocked Pipeline Stages}
\acro{MMIO}{Memory Mapped \acl{IO}}
\acro{MSB}{Most Significant Bit}

\acro{NASA}{National Aeronautics and Space Administration}
\acro{NDA}{Non Disclosure Agreement}
\acro{NSA}{National Security Agency}
\acro{NVM}{Non-Volatile Memory}

\acro{OFB}{Output Feedback Mode}
\acro{OISC}{One Instruction Set Computer}
\acro{ORAM}{Oblivious Random Access Memory}
\acro{OS}{Operating System}

\acro{PAR}{Place-and-Route}
\acro{PCB}{Printed Circuit Board}
\acro{PC}{Personal Computer}
\acro{PS}{Passive Serial}
\acro{PUF}{Physical Unclonable Function}

\acro{RISC}{Reduced Instruction Set Computer}
\acro{RNG}{Random Number Generator}
\acro{ROM}{Read-Only Memory}
\acro{ROP}{Return-oriented Programming}
\acro{RTL}{Register Transfer Level}

\acro{SCA}{Side-Channel Analysis}
\acro{SSC}{Strongly Connected Component}
\acro{SHA}{Secure Hash Algorithm}
\acro{SNR}{Signal-to-Noise Ratio}
\acro{SPA}{Simple Power Analysis}
\acro{SPI}{Serial Peripheral Interface Bus}
\acro{SRAM}{Static Random Access Memory}
\acro{SEM}{Scanning Electron Microscope}

\acro{UART}{Universal Asynchronous Receiver Transmitter}
\acro{UHF}{Ultra-High Frequency}

\acro{VHDL}{Very High Speed Integrated Circuit Hardware Description Language}

\acro{WISC}{Writeable Instruction Set Computer}

\acro{XDL}{Xilinx Description Language}
\acro{XTS}{XEX-based Tweaked-codebook with ciphertext Stealing}

\end{acronym}

%% file: sections/introduction.tex

\section{Introduction}
\label{co::sec::intro}
In an increasingly interconnected world, hardware components serve as the root of trust in virtually any computing system.
Therefore malicious hardware manipulations of mission-critical components can have serious implications, ranging from simple revenue loss over faults in critical infrastructure up to life-threatening consequences \cite{fyrbiak:2017:ivsw}.

On the one hand hardware reverse engineering is the tool-of-choice to identify such manipulations, and to check the trustworthiness of hardware in general \cite{bhunia:2014:ieee, fyrbiak:2017:ivsw}. It encompasses the detection of potentially harmful counterfeits and copyright infringements.
On the other hand, hardware reverse engineering is often used to insert hardware Trojans \cite{wallat:2017:ivsw}, which weakens the security of a system or to commit said copyright infringements by trying to counterfeit or simply copy intellectual property.

Since hardware reverse engineering is a highly complex process, semi-automated tools are desperately needed by the community \cite{wiesen:2019:aspdac}.

In the world of software reverse engineering comprehensive and expandable frameworks covering the complete workflow of binary analysis exist, e.g., IDA Pro or Ghidra.
However, for hardware reversing there is no such framework yet \cite{torrance:2009:ches} but loose collections of scripts, e.g., \cite{meade:2016:isfta}.
To the best of our knowledge, we are the first to release a fully customizable open source gate-level netlist reverse engineering framework \cite{fyrbiak:2018:tdsc} to the open-source community via GitHub\footnote{\HAL, \url{https://github.com/emsec/hal}}.

We encourage the community to conduct their own research using our framework in the field, to write and publish plugins for specialized reverse engineering tasks, and to support the development of \HAL as a whole. 
To get the interested parties started, we provide guidance for the rich feature set and its technical foundations for both \HAL users and developers.
Furthermore, we demonstrate the usage of \HAL by means of two case studies: Reverse engineering \acp{FSM}, and finding watermarks.


%% file: sections/technical_background.tex

\section{Background}
\label{cf2019::sec::background}
In the following section we provide the essential background information for the topic of chip-level reverse engineering and introduce the field of gate-level netlist reverse engineering.

\subsection{Chip-Level Reverse Engineering}

With chip-level reverse engineering, an attacker extracts a human-readable gate-level netlist from the examined \ac{IC} or \ac{FPGA}.

During this phase no functional analysis of the netlist takes place.
Only in the gate-level netlist reverse engineering step conducted later, the attacker analyzes the chip's logical functionality.

\paragraph{\textbf{\ac{FPGA} Reverse Engineering}}
Due to its volatile nature SRAM-based \acp{FPGA} are reconfigured on every boot-up by an externally stored bitstream.
The bitstream contains the configuration of the basic \ac{FPGA} elements, i.e., which Boolean function is evaluated in a \ac{LUT} and how these logical functions are connected via the routing.
For reversing an \ac{FPGA} bitstream, an attacker has to (i)~extract the bitstream from the external memory, (ii)~reverse the bitstream file format, and (iii)~convert the downloaded bitstream to a human-readable netlist.

For retrieving the bitstream the attacker can either wiretap the configuration lines on the PCB, or directly read out the flash memory. Even if the attacker encounters an encrypted bitstream the chances of recovery are high as shown in \cite{moradi:2011:ccs, moradi:2013:fpga, swierczynski:2014:trts, 10.1007/978-3-319-43283-0_5, 10.1007/978-3-642-27954-6_1}.
The bitstream file format reversing process has been described in recent papers \cite{ender:2019:aspdac, JBITS,4101017,note:2008:fpga,benz:2012:fpl,ding:2013:mm,pham:2017:date,debit,max5,prjxray}.
All these works use the correlation method.
Here, the attacker creates a basic design containing an instantiation of the examined FPGA element, e.g., a \ac{LUT}, \ac{FF}, or the routing.
In the next step, the attacker varies the elements' configuration and creates one bitstream from the basic design and one from the modified design.
The difference between both bitstreams correlates to the introduced changes in the altered design.
Using the correlation method the attacker can build a database of bitstream bits and their corresponding configuration in the netlist.
Using this database, the attacker can convert the bitstream under attack to a human-readable netlist.


\paragraph{\textbf{\ac{IC} Reverse Engineering}}
In contrast to \acp{FPGA}, reversing \acp{IC} requires several steps and is considerably more complex due to shrinking technology sizes~\cite{fyrbiak:2017:ivsw,torrance:2009:ches,
lippmann:2019:aspdac, quadir:2016:jetc}.  
The reversing steps consist of (i)~decapsulating, (ii)~delayering, (iii)~image acquisition, and (iv)~image processing in order to generate the human-readable netlist.

First, the \ac{IC} is decapsulated mostly using wet or dry chemistry to remove the organic package material or by using mechanical means.
The chemicals can fully remove the packaging, while not damaging the silicon die.
In the next step, the chip is delayered and images of each layer are acquired.
This step depends on the used manufacturing technologies, thus there exist several delayering techniques.
On today's feature sizes, the first passivation layer is often removed with dry anisotropic etching.
The next metal layers are removed via plasma etching or ion milling.
The difficulties are the over-etching -- especially of the die's edges -- or warpages due to the mechanical stress between the substrate and the metal layer.
Each of these layers are digitized via a \ac{SEM} or a \ac{FIB}.
The remaining metal layers and oxide layer is then removed with diamond suspension and dry chemistry.
Using fluoric acid the active regions of the chip are revealed.

After acquiring all images from each layer the images are stitched together.
Here, precise alignment is crucial to introduce no faulty transitions between two neighboring images.
Finally, software assisted image processing generates the human-readable netlist by identifying standard cells first and reconstructing the connections in the metal layer second.

\subsection{Gate-level Netlist Reverse Engineering}
\label{cf2019::sec::background::netlist}

A gate-level netlist is a representation of a set of logic gates from a particular gate library together with their interconnections \cite{weste:cmos:2011}.
Combinational logic is usually implemented with Boolean gates or \acfp{LUT} and multiplexers, while sequential logic is realized through \acfp{FF} or latches. All these elements are defined by the regarding gate library.
Netlists can either be represented textually via \acs{HDL} or as a graph, where the edges depict connections and the nodes represent gates.

The absence of (1)~meaningful descriptive labels, (2)~boundaries of implemented modules, and (3)~module hierarchies in flat gate-level netlists drastically complicates the process of gate-level netlist reverse engineering \cite{fyrbiak:2017:ivsw}.

However, the representation as a graph facilitates the application of graph-based algorithms, which can help identifying the control logic or restoring certain module boundaries and hierarchies.
A further approach consists in the detection of unique (logical) structures in the netlist.

%% file: sections/hal.tex
\section{HAL -- The Hardware Analyzer}
\label{cf2019::sec::hal}
\HAL aids analysts with a rich feature set to facilitate explorative functionality recovery of gate-level netlists in a semi-automated fashion.
To this end \HAL processes netlists in its own graph-based representation (cf. \Section{cf2019::sec::background::netlist}).
Note that \HAL itself is gate-library agnostic, hence it can be used to analyze netlists of \acp{ASIC} as well as \acp{FPGA}.

\begin{figure*}[!htb]
	\centering
	\includegraphics[width=0.7\linewidth]{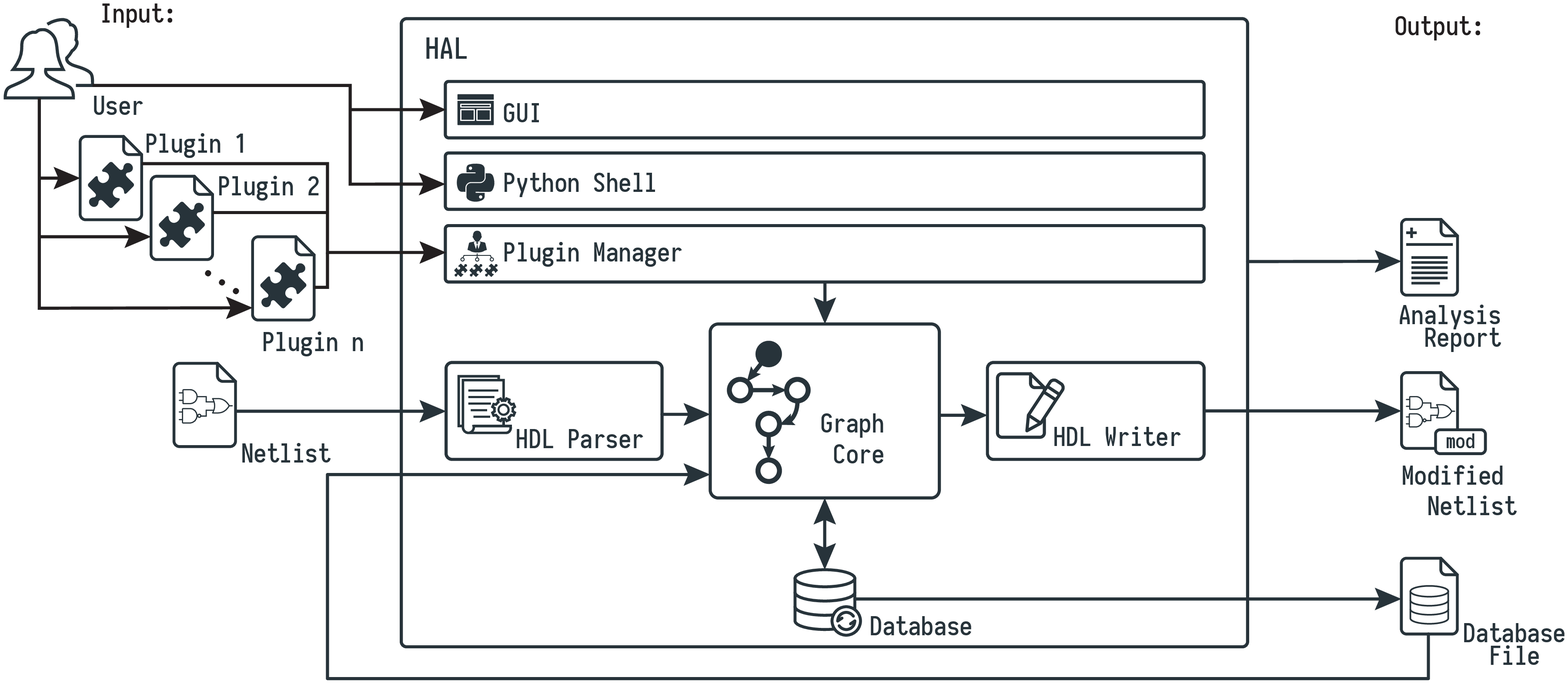}
	\caption{Overview of the original \HAL architecture from \cite{fyrbiak:2018:tdsc} (modified)}
	\label{cf2019::fig::hal_structure}
\end{figure*}

\Figure{cf2019::fig::hal_structure} shows a high-level overview of the main workflow and core features of \HAL.
An analyst can either parse a new netlist into \HAL or continue previous work by loading a snapshot file.
After \HAL has loaded the netlist, the analyst can use \HAL's graph core to freely traverse or even manipulate the netlist.
This can be done in an explorative manner in the GUI, either via the Python shell or by direct interaction with the graph view.
To perform time-critical algorithms without the performance penalties of Python, custom C++ plugins can be used, even via the Python shell.
All actions performed are documented in log files and plugins can access their own logging channels to allow for straightforward report filtering.
Changes to the graph, regardless of the origin, are directly reflected in the GUI elements, allowing the analyst to not only logically but also visually partition the netlist.
At any point in time, the analyst can create a snapshot of the current graph representation, that can be used to resume analysis later or revert to an earlier state.

Note that the GUI is entirely optional.
Hence, \HAL can be executed as a stand-alone command line tool, offering its full range of features except for visualization.
After performing modifications, the analyst can choose to write the netlist back into a an \acs{HDL} format, resulting in a synthesizable gate-level netlist.

\paragraph{\textbf{Open Source Release}} Due to the growing demand from the scientific community we decided to publicly release \HAL.
The source code is available on GitHub (cf. \Section{co::sec::intro}) under the open-source MIT license. We hope that \HAL will be of use to the research community and encourage interested developers to contribute to the project via GitHub. We support both Linux and macOS as the \ac{OS}.

\subsection*{Technical Foundation}

Throughout the development of \HAL various aspects regarding the performance, usability, and modular expandability had to be considered. The following section highlights the emerged issues and presents our solutions.

\paragraph{\textbf{The Core System}}

Since complex gate-level netlists are composed of several thousands up to billions of gates and interconnections, performance posed an urgent issue from the very beginning of the development of \HAL.

\begin{figure}[H]
	\centering
	\includegraphics[width=0.61\linewidth]{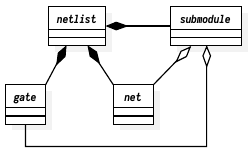}
	\caption{Simplified \HAL netlist library class diagram}
	\label{co::fig::hal_data_structure}

\end{figure}

Therefore, the low-level programming language C++17 was chosen to implement the underlying core system.
Here the netlist library, which represents the data structure for all elements of a netlist, constitutes the crucial component. The class diagram in \Figure{co::fig::hal_data_structure} depicts the relationship between the core classes.

In contrast to off-the-shelf graph libraries, the netlist library has the following distinct properties, which are specifically designed for netlist processing.
\begin{description}
	\item[Gate] Each gate object has a gate type (e.g., NAND, NOR,~\dots) dynamically assigned based on the underlying gate library while parsing the netlist. 
        Additional information, e.g., LUT configuration strings, FF init values, etc., are stored directly in the gate object.
	\item[Net] In contrast to classical edges with a single source and sink a net in our library allows to have multiple sinks.
	\item[Submodule] To add hierarchy information during the reverse engineering process additional submodules can be defined. Each submodule lists the gates and nets belonging to the submodule.
\end{description}
\begin{figure*}[!htb]
	\centering
	\includegraphics[width=\linewidth]{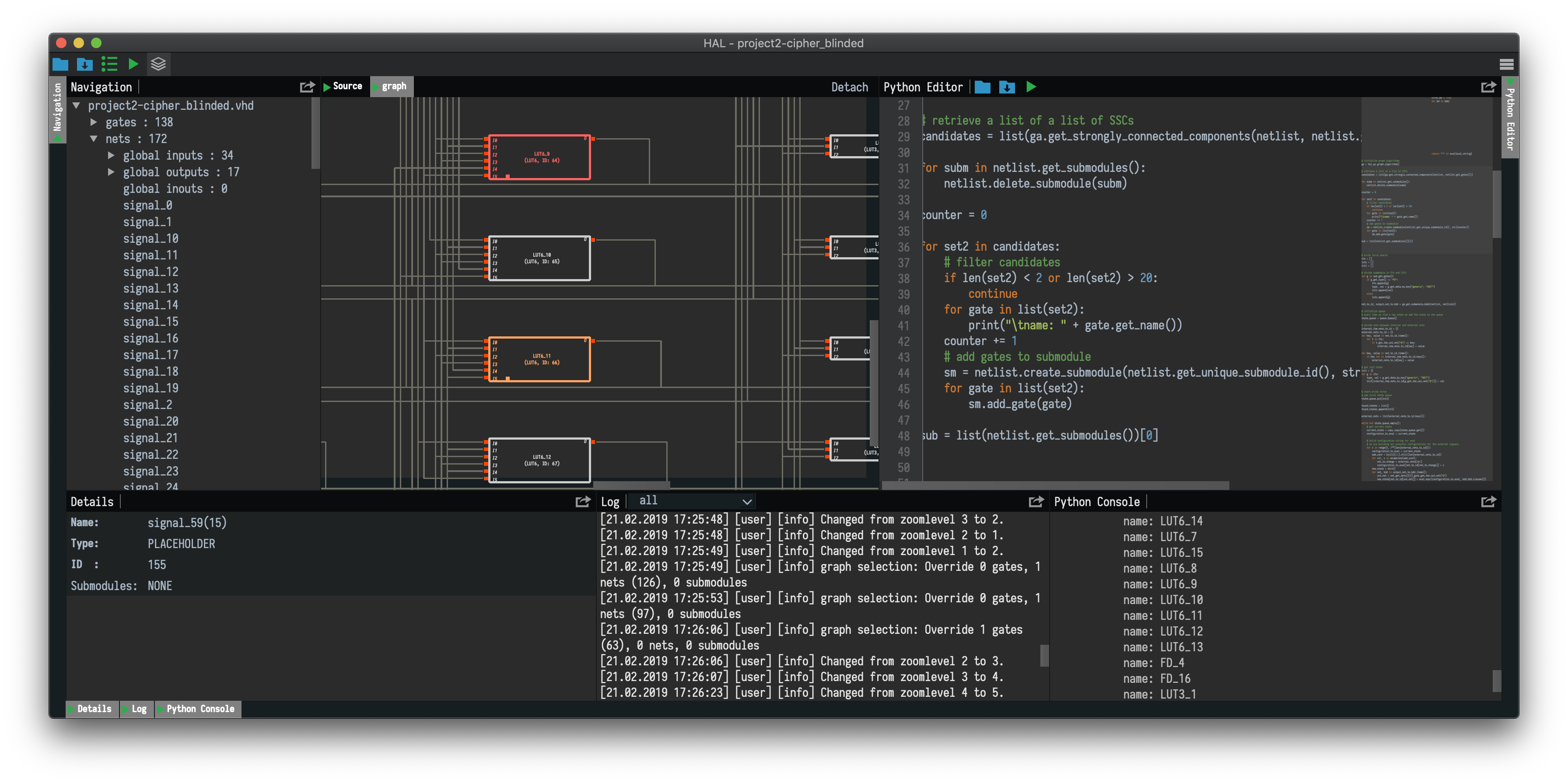}
	\caption{Screenshot of the \HAL \acl{GUI} during the Reverse Engineering Process}
	\label{co::fig::hal_gui}
\end{figure*}
In addition to the core module, the netlist library introduces an event system allowing other components to be notified when the underlying data model changes its information. This is specifically necessary for interactive components like the \ac{GUI}.

\paragraph{\textbf{The Plugin System}}

Due to the collision between the requirement of working on netlist reversing projects under an \ac{NDA} and the goal to provide a collaborative open-source framework to the community, we decided to introduce a plugin system. It allows leaving the core parts of the system public while project-specific elements can be placed in a plugin without the necessity to publish them.
The plugin system is realized through C++ dynamic libraries which are loaded on demand by the core.
This allows for straightforward parallelization of computation-intense tasks, for example via OpenMP.

As an example, we provide a plugin for dynamic graph analyses called \texttt{graph-algorithm} which allows further processing of a netlist in \HAL using the Boost Graph library\footnote{Boost Graph Library, \url{https://www.boost.org/doc/libs/1_66_0/libs/graph/doc/index.html}}. 

\paragraph{\textbf{The Graphical User Interface}}

While algorithmic analyses of netlists are a powerful tool, there are cases where the visual inspection of a design is necessary.
Therefore, our 
Qt5\footnote{Qt5, \url{https://www.qt.io/}}-based \ac{GUI} provides a performant graphical representation even for large netlists and enables interactive navigation with mouse and keyboard through the graph.
At the same time, a navigation pane ensures that the user always maintains an overview, while additional information about the selected netlist component is displayed in a detail pane.

A further GUI-feature is the color-based submodule highlighting, which supports users in the process of understanding the inner workings of a design in combination with self-developed plugins for algorithmic analyses.

Altogether, the \ac{GUI} facilitates the reverse engineers' task to process the given information and to make sense of a formerly unknown design \cite{wiesen:2018:tale}.

\paragraph{\textbf{Python Integration}}

To lower the barrier of entry for new \HAL users and developers, we embedded a Python shell into the \ac{GUI}.
The Python shell provides an efficient and intuitive approach to interact with a netlist; whereas the development of custom C++ plugins offers full flexibility, but requires more experience.

From a technical perspective, we employed pybind11\footnote{Pybind11, \url{https://github.com/pybind/pybind11}}
to map the C++ API to the smaller and simpler Python API.
All function calls from the fully-featured Python~3.7.2 interpreter to the core are handled by the C++ back-end to preserve its performance advantages.

%% file: sections/case-studies.tex
\section{Case Studies}
\label{cf2019::sec::case-studies}
In the following we present two case studies demonstrating the capabilities and flexibility of \HAL in reverse engineering gate-level netlists.

\subsection{Reverse Engineering Finite States Machines}
\label{cf2019::sec::case-studies::fsm}
Since an \ac{FSM} controls almost every hardware design it presents a promising attack target for reverse engineers. Fyrbiak \etal \cite{fyrbiak:2018:ches} proposed a method for finding \ac{FSM} circuits in a netlist as well as a way of retrieving the corresponding state-graph. For extracting \acp{FSM} from a netlist a plugin in \HAL has been created.

\begin{figure}[!htb]
	\centering
	\includegraphics[width=\linewidth]{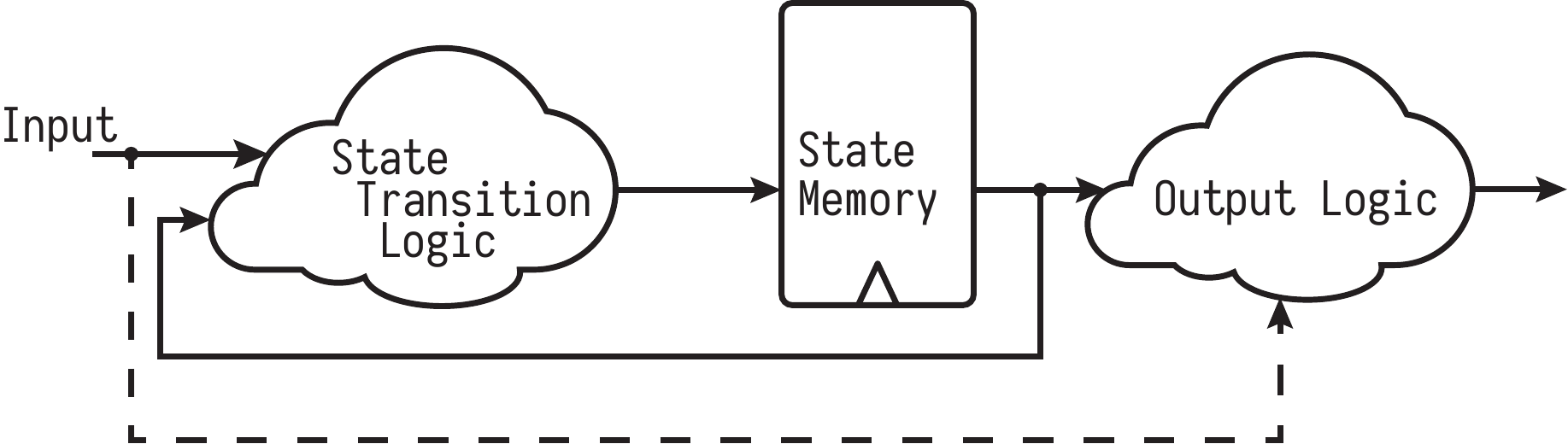}
	\caption{Block diagram of a hardware FSM (dashed line in the case of a Mealy machine) from \cite{fyrbiak:2018:tdsc}}
	\label{cf2019::fig::fsm_structure}
\end{figure}

From a mathematical perspective \acp{FSM} are equivalent to a \ac{SSC} (see \Figure{cf2019::fig::fsm_structure}). There are several algorithms from graph theory that can be used to identify \acp{SSC}. In \HAL we implemented Tarjan's Algorithm \cite{tarjan:1971:swat} to identify \acp{SSC} in our own \texttt{graph\_algorithm} plugin. Once the plugin has been loaded, the corresponding functions can operate on the netlist and report back the results to the main program, where they can be used for further analysis.

After the \textit{correct} \ac{FSM} circuit has been identified, the state graph can be retrieved by analyzing the Boolean logic of the state transition logic. We conducted our analysis for \ac{FPGA} netlists, which incorporate mostly \acp{LUT} to realize the logic. Using the functionalities provided by \texttt{gate-decorators}, \HAL offers the possibility to generate a \ac{BDD} representing the logic expression for one or multiple \ac{LUT}-gates. A \texttt{gate-decorator} extends the already available functions for gates provided by the \texttt{gate} class. \texttt{Gate-decorators} are specific for every gate-library. This means, for porting said method to extract \acp{FSM} from \ac{ASIC} netlists, one has to provide the specific definitions of the gate-library in order to generate the corresponding \ac{BDD}.
With the help of the generated \acp{BDD}, we can brute-force all reachable states for reasonably bounded \acp{FSM} without the need for further libraries.
Of course, computation time grows quickly with the complexity of the feedback logic and the \ac{FSM} state register's size.
In the end, we output the state graph as a GRAPHVIZ \texttt{.dot} file.

\paragraph{\textbf{Circumventing FSM Obfuscation}}
Obfuscation describes the transformation, which obstructs high-level information without changing functionality while increasing the complexity of the reverse engineering process in mind.

Fyrbiak \etal \cite{fyrbiak:2018:ches} also described means of attacking several well-known obfuscation schemes operating on the \ac{FSM}-level \cite{chakraborty:2009:tcad, dofe:2017:tcad, alkabani:2007:usenix}. Since \HAL offers netlist manipulation techniques, we can efficiently implement a plugin to circumvent, remove, or disable obfuscation techniques.

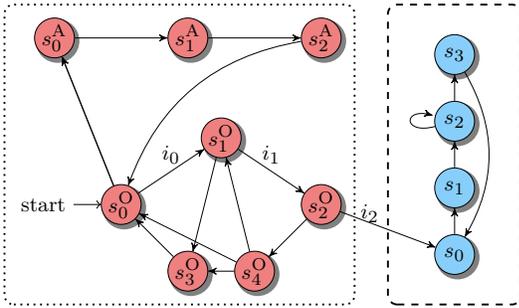
\begin{figure}[htb]
    \centering
    \resizebox{0.85\linewidth}{!}{%
        \begin{tikzpicture}
        \tikzstyle{every state}=[fill=white,draw=black,text=black,inner sep=0pt,minimum size=17pt,circular drop shadow]
        
        \draw [help lines,red] (0,0) (8,5);
        
        \node[initial,state,fill=lightcoral] (sO0) at (2,2) {$s^\text{O}_0$};
        \node[state,fill=lightcoral] (sO1) at (3.5,3) {$s^\text{O}_1$};
        \node[state,fill=lightcoral] (sO2) at (5,2) {$s^\text{O}_2$};
        \node[state,fill=lightcoral] (sO3) at (3,1) {$s^\text{O}_3$};
        \node[state,fill=lightcoral] (sO4) at (4,1) {$s^\text{O}_4$};
        \node[state,fill=lightcoral] (sA0) at (1,4.5) {$s^\text{A}_0$};
        \node[state,fill=lightcoral] (sA1) at (3,4.5) {$s^\text{A}_1$};
        \node[state,fill=lightcoral] (sA2) at (5,4.5) {$s^\text{A}_2$};
        \node[state,fill=lightskyblue] (s0) at (7,1.25) {$s_0$};
        \node[state,fill=lightskyblue] (s1) at (7,2.25) {$s_1$};
        \node[state,fill=lightskyblue] (s2) at (7,3.25) {$s_2$};
        \node[state,fill=lightskyblue] (s3) at (7,4.25) {$s_3$};
        
        \path[->,>=stealth']
        (sO0) edge [above] node {$i_0$} (sO1)
        (sO0) edge (sA0)
        (sO1) edge [above] node {$i_1$} (sO2)
        (sO1) edge (sO3)
        (sO2) edge (sO4)
        (sO2) edge [anchor=center, above] node {$i_2$\ \ \ \ \ \ } (s0)
        (sO3) edge (sO0)
        (sO4) edge (sO3)
        (sO4) edge (sO0)
        (sO4) edge (sO1)
        (sO0) edge (sA0)
        (sA0) edge (sA1)
        (sA1) edge (sA2)
        (sA2) edge [bend right] (sO0)
        (s0) edge (s1)
        (s1) edge (s2)
        (s2) edge (s3)
        (s2) edge [loop left] (s2)
        (s3) edge [bend left] (s0);
        
        \draw [rounded corners, thick, dotted] (0.25,0.5) -- (5.5,0.5) -- (5.5,5) -- (0.25,5) -- cycle;
        \draw [rounded corners, thick, dashed] (6,0.5) -- (8,0.5) -- (8,5) -- (6,5) --cycle;
        
        \end{tikzpicture}
    }
    \caption{Obfuscated \ac{FSM} using HARPOON \cite{fyrbiak:2018:ches, chakraborty:2009:tcad}}
    \label{cf2019::fig::obfsucated_fsm}
\end{figure}

 One of the most popular obfuscation schemes is HARPOON \cite{chakraborty:2009:tcad}. The basic idea of HARPOON - see \Figure{cf2019::fig::obfsucated_fsm} - envisages a designer inserting a second \ac{FSM} (highlighted in red) to protect the original \ac{FSM} (highlighted in blue). The inserted \ac{FSM} part has to be traversed in a certain way, using a specific input sequence called the enabling key. Every other input sequence than the enabling key will not lead to the original \ac{FSM}, thus rendering the hardware design unusable for unauthorized parties \cite{wiesen:2019:aspdac}.

  Fyrbiak \etal \cite{fyrbiak:2018:ches} proposed a general attack idea to (i) find a HARPOON key and (ii) remove the HARPOON key from the netlist. We introduce and use these attack ideas to present various features of \HAL, e.g., the netlist manipulation, and plugin features. First, with the brute-force attack described in \Section{cf2019::sec::case-studies::fsm}, we executed the \ac{FSM} detection plugin and read the HARPOON enabling key from the extracted state graph.

Second, we completely changed the behavior of the state machine and generate a manipulated netlist. For that, we use the netlist manipulation feature of \HAL.
Changing the initial value of the Flip-Flops from the state memory to the values of the initial state of the original \ac{FSM} results in omitting the obfuscated part. This removing of the obfuscation \acp{FSM} is possible as we know the initial states from the first attack step.
The manipulated netlist can be written to either Verilog or \ac{VHDL}. In case of an \acp{FPGA} netlist, a new bitstream can be generated with the corresponding vendor tools. \HAL even allows manipulating the transition logic and thus manipulating the behavior of the state machine.

 \subsection{Finding Watermarks}
 In the context of hardware design, a watermark is a secret or hidden \textit{message} inside a circuit that enables the owner of the design to identify his work. It is usually used in the context of IP-infringement to identify intellectual property.

   \begin{figure}[!htb]
 	\centering
 	\resizebox{0.85\linewidth}{!}{%
 		\begin{tikzpicture}[font=\sffamily,>=triangle 45]


 		\node [shape=lut] (LUT1) at (0,0) {\small\texttt{LUT3}};

 		\draw [<-] (LUT1.IONE) -- +(-0.5,0) node [anchor=east] {a};
 		\draw [<-] (LUT1.ITWO) -- +(-0.5,0) node [anchor=east] {b};
 		\draw [<-] (LUT1.ITHREE) -- +(-0.5,0) node [anchor=east] {GND};

 		\draw [->] (LUT1.Q) -- +(0.5,0) node [anchor=west] {c};

 		\node (table) at ([xshift=4cm]LUT1) {
 			\resizebox{!}{!}{
 				\begin{tabular}{@{}ccc|c@{}}
 				\toprule
 				\textbf{I2}				& \textbf{I1}			& \textbf{I0}			& \textbf{O}		\\
 				\midrule
 				0 				& 0 		   & 0 			   & 1 		\\
 				0 				& 0 		   & 1 			   & 0 		\\
 				0 				& 1 		   & 0 			   & 1 		\\
 				0 				& 1 		   & 1 			   & 0 		\\
 				1 				& 0 		   & 0 			   & Y 		\\
 				1 				& 0 		   & 1 			   & Y 		\\
 				1 				& 1 		   & 0 			   & Y 		\\
 				1 				& 1 		   & 1 			   & Y 		\\
 				\bottomrule
 				\end{tabular}%
 			}
 		};
 		\end{tikzpicture}%
 	}
 	\caption{Overview of the watermarking scheme}
 	\label{cf2019::case_studies::watermarking}
 \end{figure}
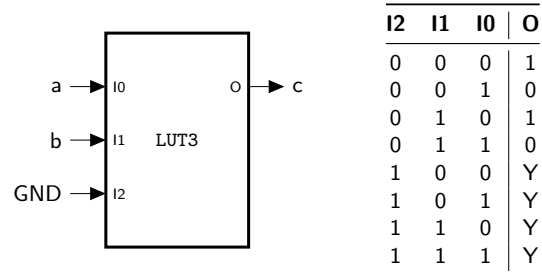

 Wallat \etal \cite{wallat:2017:ivsw} proposed ways for identifying and removing a watermark scheme presented by Schmid \etal \cite{schmid:2008:fpt} for \acp{FPGA}.

The scheme makes use of the fact that sometimes not all inputs of a \ac{LUT} are in use. If a \ac{LUT} has an unused input, it is usually being connected to either \texttt{GND} or \texttt{VCC} -- respectively logic '0' or '1' . If a \ac{LUT} has an input connected to \texttt{GND} or \texttt{VCC} it results in unreachable entries in the truth table -- see \Figure{cf2019::case_studies::watermarking} -- the entries marked with \texttt{Y} are the entries that cannot be reached. Their watermarking scheme inserts a unique sequence into these unreachable entries to uniquely mark a design.

 We used \HAL to identify \acp{LUT} that were used for the watermarking by analyzing the \ac{LUT} content of all \acp{LUT} that have \texttt{GND} or \texttt{VCC} connected. For each of these \acp{LUT} the \ac{LUT} content is analyzed for entries that where not set to '0', when they cannot occur. This way the watermarking can easily be identified. Furthermore we removed the watermarking using \HAL, by manipulating the \ac{LUT}'s content using the netlist manipulation feature.


%% file: sections/conclusion.tex

\section{Conclusion}
Hardware reverse engineering as the tool-of-choice to examine hardware designs for their functionality and potential manipulations, or to detect product counterfeits. Due to the lack of publicly available and fully-customizable frameworks assisting the gate-level netlist reversing process we present our gate-level netlist reverse engineering framework \HAL.
Furthermore we present its rich feature set providing visual and algorithmic access to gate level netlists, as well as its technical foundations to get potential users started.
In an effort to involve the open-source community into the development, we release the \HAL source code on \url{https://github.com/emsec/hal} under the MIT open-source license.

A main feature of \HAL is the representation of the netlist as a graph which enables further graph-based analyses.
In two case studies we demonstrated the manifold capabilities of \HAL:
First, we illustrate the creation of plugins to simplify the netlist reverse engineering process in a practical context. 
Second, we demonstrated how the \texttt{graph\_algorithm} plugin can be applied to identify structures and modules within the flat netlist.
In the end, the powerful manipulation feature shows how the behavior of a netlist can be changed to circumvent real-world obfuscation techniques.
\enlargethispage{\baselineskip}